# Chromatic effects in beam wander correction for free-space quantum communications

Alberto Carrasco-Casado, Natalia Denisenko, Veronica Fernandez

Spanish National Research Council (CSIC)—Institute of Physical and Information Technologies (ITEFI), Serrano 144, Madrid 28006, Spain
Corresponding author: veronica.fernandez@iec.csic.es

*Abstract*—Beam wander caused by atmospheric turbulence can significantly degrade the performance of horizontal free-space quantum communication links. Classical beam wander correction techniques cannot be applied due to the stronger requirements of transmitting single photons. One strategy to overcome this limitation consists in using a separate wavelength from that of the quantum signal to analyze the beam wander and use this information for its correction. For this strategy to work adequately, both wavelengths should be affected equally by atmospheric turbulence, i.e. no chromatic effects should be originated from beam wander. In this letter, a series of experiments are performed to prove that this is the case for $\lambda \sim 850$ nm as the quantum signal and $\lambda \sim 1550$ nm as the tracking signal of a quantum communication system.

## I. INTRODUCTION

Quantum Key Distribution (QKD) [1] is the most technologically developed branch of quantum communications and consists in the establishment of a secure key between two parts. The secrecy of this distribution is based on fundamental properties of quantum physics rather than assumptions on computational complexity, as it is the case in classical cryptography. The transmission channel has traditionally been optical fiber or free space. The latter has experienced a strong development in the last few years, reaching a link distance of 144 km [2, 3], due to its potential to extend the range of quantum links. Initially, it was primarily aimed towards satellite communication [4] and the main efforts have been focused in achieving long distances to prove its feasibility. However, short/medium distance (inter-city range) horizontal links in urban areas are also of considerable interest [5-7], since they have the potential of alleviating connectivity bottlenecks in metropolitan quantum communication networks. Moreover, their flexibility of installation and portability makes them an interesting alternative when the installation of optical fiber is not viable or in the presence of an operational failure in the network (natural catastrophes, for example).

## II. BEAM WANDER CORRECTION IN QUANTUM COMMUNICATIONS

The capabilities of urban free-space communication links can be optimized including automatic tracking systems, which correct for beam wander, i.e. fast deflections in the received beam caused by atmospheric turbulence. Although beam wander correction is a well-known technique in classical free-space systems, there are important differences to consider when applied to quantum communications. In classical links, a fraction of the received optical signal is used to analyze beam wander and correct for the deviations of the main signal. Conversely, in quantum communications, where single photons are transmitted, this technique cannot be applied in the same way, since it would cause additional losses in the system.

In a previous work, the authors proposed a beam wander correction system for quantum communications [8]. This system uses a separate wavelength for analyzing the beam wander of the received beam, referred to as the tracking channel, and another one for carrying the useful data of the quantum communication link, referred as the quantum channel. Since an additional channel is needed to perform the timing synchronization between transmitter and receiver, it was proposed to use the same wavelength for synchronization and tracking to simplify the setup.

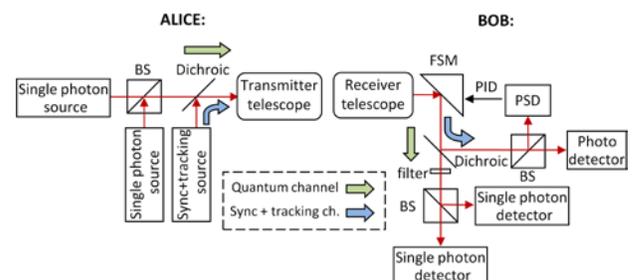

Fig. 1. Diagram of beam wander correction for the B92 protocol.

Fig. 1 shows a simplified diagram of this system for the B92 protocol. First, Alice transmits three aligned laser beams (two beams for the quantum channel and one beam for the synchronization and tracking channel) through a telescope, and Bob gathers their photons with another telescope. Then, quantum and sync+tracking channels are spectrally discriminated through a dichroic mirror. The quantum channel is directed to single-photon detectors through the appropriate polarization control (not shown in Fig.1). The sync+tracking is split in two: one fraction goes towards a photodetector (to perform timing synchronization) and another to a PSD, a Position Sensitive Detector (tracking channel). The signal detected in the PSD is then used to control a FSM (Fast-Steering Mirror), which corrects the received beam deflections through a PID (Positive Integral Derivative) control. It should be stressed that the goal is the correction of the beam deflections that the quantum channel experiences as it propagates through the transmission channel, albeit its own beam deflections are not measured, but the tracking channel deflections instead. Therefore, for this strategy to work adequately, all beams must suffer the same atmospheric perturbations.

For a number of reasons (being the main one the suitability of photodetectors), free-space quantum communication links are usually implemented using a wavelength around 850 nm, as it is also the case in the proposed system. As for the tracking and synchronization channel, $\lambda \sim 1550$ nm is employed, for having several advantages such as eye safety, lower atmospheric attenuation and off-the-shelf availability.

## III. WAVELENGTH DEPENDENCE OF BEAM WANDER

The strategy of beam wander correction mentioned before consists in monitoring in the receiver the beam deflections suffered by the 1550-nm beam to correct blindly for the deflections of the 850-nm beam. If the beam wander caused by atmospheric turbulence has a different effect depending on the wavelength, then this strategy cannot be applied. Therefore, the goal of this study is to study whether turbulence-related beam wander show a chromatic or an achromatic behavior. In [9], an experiment was carried out to compare the effect of atmospheric turbulence over these two wavelengths, showing a good correlation between them, although only signal intensities were measured. However, in a beam wander correction system, beam position must also be measured, as it gives a more detailed description of the deviations due to atmospheric turbulence than signal intensity.

An experiment was performed to measure the instantaneous centroid position of two collimated and aligned laser beams at $\lambda \sim 850$ nm and $\lambda \sim 1550$ nm after been affected by the same turbulence. The turbulence was emulated with an air heater placed near the transmitter lasers, which are located 5 meters away from the receiver. This is a common practice to simulate longer paths affected by atmospheric-like turbulence when performing laboratory experiments [10].

Fig. 2 shows the experimental setup used for these measurements, where both signals at $\lambda \sim 850$ nm and $\lambda \sim 1550$ nm are collimated to attain an output beam diameter of 7 mm and then combined using a beamsplitter. The turbulence is applied next to the transmitter, and after a propagation of 5 meters, both beams are separated using a dichroic mirror and focused towards two position sensitive detectors. Lastly, both signals are detected in a synchronous mode to perform the correlation analysis. A fast-steering mirror was used to calibrate the receiver system for each beam to show the same behavior under the same perturbations. This ensures that any difference between both wavelengths will be caused by chromatic effects from the turbulence.

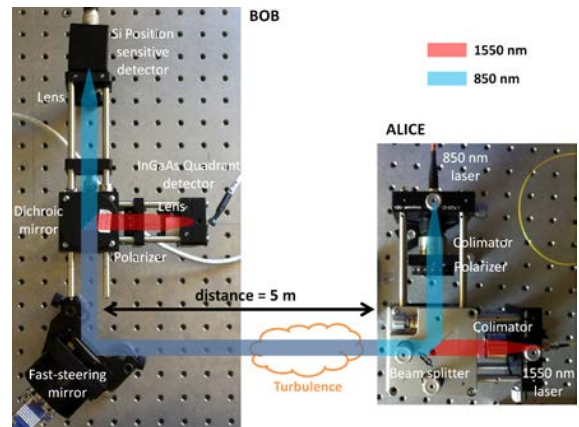

Fig. 2. Experimental setup for measuring chromatism in beam wander.

The detector used for measuring the 1550-nm path was an InGaAs quadrant detector (IGA-030-QD by EOS Systems), with a diameter of 3 mm, and that used for $\lambda \sim 850$ nm was a Silicon lateral effect detector (C10443-02 by Hamamatsu), with a diameter of 10 mm. Their signals were conditioned with transimpedance amplifiers and X-Y position calculation circuits, and processed with the USB6343 data acquisition board and a LabView by National Instruments.

It is important to note that both beams have to be collimated with the exact same diameter and have to be perfectly aligned within the same optical axis. Otherwise, each beam would propagate through a slightly different optical path and would be affected by beam wander in a different way, regardless of the chromatic or achromatic behavior of atmospheric turbulence.

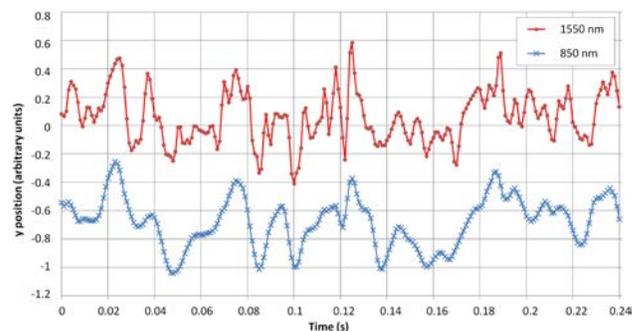

Fig. 3. Centroid position of two beams of $\lambda \sim 850$ nm and $\lambda \sim 1550$ nm experiencing beam wander.

Fig. 3 shows a sample result of this experiment during a perturbation of 240 ms over the Y axis after calibrating and synchronizing both channels. At first glance, it can be seen that both signals follow the same movements, except for the higher frequencies. This can be explained by taking the bandwidth of both detection systems into consideration. The 1550-nm channel was characterized by having a 3-dB bandwidth of 16 kHz, exactly two orders of magnitude above the bandwidth of the 850-nm channel. This explains the high-frequency differences between both channels, as the 850-nm channel cannot follow the faster response of the simulated turbulence.

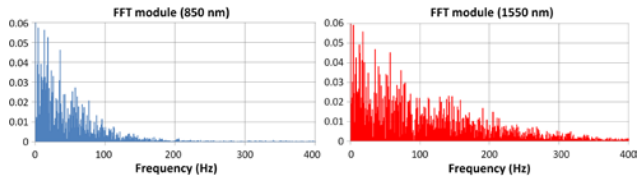

Fig. 4. FFT module of beam wander in 850 nm and 1550 nm.

In fig. 4, a FFT module of the detected signals from both channels is shown. Since all the frequency components of the 1550-nm signal remain below the channel bandwidth, this is in fact a frequency analysis of the simulated turbulence. Therefore, it can be concluded that the spectrum of the 850-nm signal is low-pass filtered by the receiver's electronics, which explains that the 850-nm signal cannot detect the highest spectral components observed in the 1550-nm signal.

To mitigate the effect of the bandwidth difference between both channels, a moving average was applied to filter the higher frequency components of the turbulence detected in the 1550-nm channel. Finally, the Pearson coefficient was calculated to compute the correlation between both signals. This coefficient offers the linear relation between two random variables with a covariance-like result but isolating the scale of the measurement [11], which is especially useful when using different types of detector technologies.

After performing these calculations, a 95% correlation was computed over a sufficiently long time (several seconds) of beam wander perturbation. Less than 80% correlation was obtained when not filtering to eliminate the highest frequencies, which proves that the differences come from the limited bandwidth of the 850-nm channel. Nonetheless, most of these highest frequencies are caused by the air heater and are not present in atmospheric turbulence. This was concluded after characterizing a 30-m propagation path affected by atmospheric turbulence using the 1550-nm system.

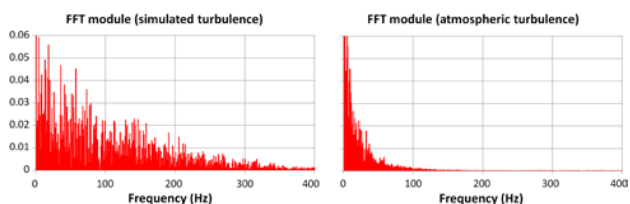

Fig. 5. FFT module of simulated turbulence at the laboratory and real atmospheric turbulence in a 30-m link.

The refractive index structure parameter $C_n^2$ was calculated from measurements of the beam centroid position for a 30-m free-space link, and estimated to be in the order of $C_n^2 \simeq 10^{-13}$ m$^{-2/3}$, which is considered a strong turbulence regime. The FFT module is shown in fig. 5, where it can be seen that natural turbulence has a much lower bandwidth than the simulated turbulence. In this experiment, the spectral components of atmospheric turbulence almost completely disappear above 200 Hz, being most of them below 100 Hz, which is less than the bandwidth of the 850-nm channel, validating the low-pass filtering that was performed. The result of this last experiment agrees well with measurements taken in longer propagation paths [12].

## IV. CONCLUSION

A proposed strategy to correct beam wander caused by atmospheric turbulence in horizontal free-space quantum communication links involves the use of a separate wavelength from the quantum signal for correcting beam wander. This has the advantage of avoiding additional losses incurred by using part of the quantum signal for the correction, as is common practice in conventional free-space optical communications. However, this strategy requires that both wavelengths are equally affected by atmospheric turbulence. The goal of this letter was to prove that this is exactly the case, and a series of experiments were performed using $\lambda \sim 850$ nm and $\lambda \sim 1550$ nm, which are the most common wavelengths chosen for free-space quantum communication links. It was showed that both signals behave in the same way under the same perturbation, with no chromatic effects related to beam wander, thus validating the proposed correction technique.

### ACKNOWLEDGMENT

We would like to thank the project TEC2012-35673 from the Ministerio de Economía y Competitividad (Spain).

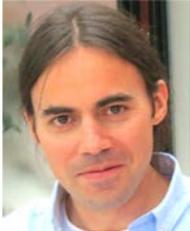

**Alberto Carrasco-Casado** received his B.S. degree in telecommunication engineering from Málaga University (Spain) in 2005. He received a M.S. degree in telecommunication engineering in 2008 and a M.S. degree in space research in 2009, both from Alcalá University, Madrid (Spain). He received his Ph.D. in electric, electronic and automation engineering from Carlos III University of Madrid and Spanish National Research Council in 2015, and then he joined the Space Communication Systems Laboratory, at NICT (National Institute of Information and Communications Technology) in Tokyo (Japan). His research interests focus on free-space optics, satellite communications, space engineering and quantum cryptography.

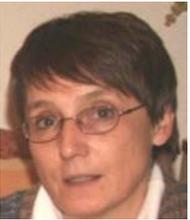

**Natalia Denisenko** received her B.S. degree in electrical and electronic engineering from Moscow Power Engineering Institute in 1975. She worked at Laser Physics Laboratory (Paris University, France) between 1975 and 1977, at Materials Physics Institute (Spanish National Research Council, Madrid, Spain) between 1979 and 1982 and at Ultrasonic and Acoustic Technologies Laboratory (National Research Council, Rome, Italy) between 1982 and 1985. In 1986, she joined the Spanish National Research Council (Madrid, Spain).

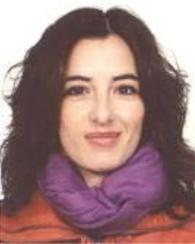

**Veronica Fernandez** received the B.Sc. degree (hons.) in physics with electronics from the University of Seville (Spain) in 2002 and the Ph.D. degree in physics from Heriot-Watt University, Edinburgh (U.K.), in 2006. In 2007, she joined the Cryptography and Information Security Group at the Spanish National Research Council (CSIC), where she was granted a tenured scientist position in 2009. Her research areas include fiber and free-space based quantum cryptography and active correction of atmospheric turbulence for free-space quantum communications.